\documentclass[preprints,article,submit,oneauthors,xelatex]{Definitions/mdpi} 

\firstpage{1} 
\pubvolume{1}
\issuenum{1}
\articlenumber{0}
\pubyear{2022}
\copyrightyear{2020}
\datereceived{} 
\dateaccepted{} 
\datepublished{} 
\hreflink{https://doi.org/}

\usepackage{dcolumn}
\usepackage{amsmath}

\usepackage{sidecap}
\usepackage{dsfont}

\usepackage{mathrsfs}
\usepackage{amsfonts}
\usepackage{indentfirst}
\usepackage{bm}
\usepackage{mathtools}
\usepackage{soul}
 \soulregister{\cite}7

\newcommand{\be}{\begin{equation}}
\newcommand{\ee}{\end{equation}}
\newcommand{\bey}{\begin{eqnarray}}
\newcommand{\eey}{\end{eqnarray}}
\newcommand{\bw}{\begin{widetext}}
\newcommand{\ew}{\end{widetext}}

\newcommand{\ra}{\rangle}
\newcommand{\la}{\langle}

\newcommand{\ba}{\begin{array}}
\newcommand{\ea}{\end{array}}
\newcommand{\bi}{\begin{itemize}}
\newcommand{\ei}{\end{itemize}}
\newcommand{\bem}{\begin{enumerate}}
\newcommand{\eem}{\end{enumerate}}

\Title{Quantum Chaos in the Extended Dicke Model}

\TitleCitation

\Author{Qian Wang $^{2,1}$}

\address{
$^{1}$ \quad CAMTP-Center for Applied Mathematics and Theoretical Physics, 
University of Maribor, SI-2000 Maribor, Slovenia \\
$^{2}$ \quad Department of Physics, Zhejiang Normal University, Jinhua 321004, China}

\AuthorNames{Qian Wang}

\AuthorCitation{Wang, Q.}

\abstract{
We systematically study the chaotic signatures in a quantum many-body system
consisting of an ensemble of
interacting two-level atoms coupled to a single-mode bosonic field, 
the so-called extended Dicke model. 
The presence of the atom-atom interaction also leads us to explore how the atomic interaction affects
the chaotic characters of the model.
By analyzing the energy spectral statistics and the structure of eigenstates, we
reveal the quantum signatures of chaos in the model 
and discuss the effect of the atomic interaction.  
We also investigate the dependence of the boundary of chaos extracted 
from both eigenvalue-based and eigenstate-based indicators on the atomic interaction. 
We show that the impact of the atomic interaction on the spectral statistics is stronger than on
the structure of eigenstates.
Qualitatively, the integrablity-to-chaos transition found in the Dicke model is amplified when the
interatomic interaction in the extended Dicke model is switched on.}

 
\begin{document}

\section{Introduction}

In recent years, the study of quantum chaos in many-body systems has attracted much attention,
both theoretically and experimentally, in different fields of physics, such as statistical physics
\cite{Altland2012,LucaD2016,Borgonovi2016,Nandkishore2015,Deutsch2018}, 
condensed matter physics 
\cite{ChanA2018,March2018,Friedman2019,Ray2020,Garttner2020,
Kobrin2021,Fogarty2021,Wittmann2022}, 
high energy physics 
\cite{Maldacena2016,Stanford2016,Magan2018,Jahnke2019,Tibra2020,Rabinovici2021},
as well as quantum information science
\cite{Schack1996,Vidmar2017,Piga2019,Bertini2019,Lerose2020,Lantagne2020,Anand2021}.
To some extent, this great interest in quantum many-body chaos is due to the close connections
of chaos to several fundamental questions that arise in current theoretical and experimental
studies.
Although a full understanding of quantum many-body chaos is still lacking, much progress has been
achieved.
It is known that chaos in interacting quantum many-body systems 
can lead to thermalization \cite{Altland2012,LucaD2016,Borgonovi2016},
the fast scrambling of quantum information \cite{Maldacena2016,Hosur2016,Chenu2019,Prakash2020}, 
an exponential growth of quantum complexities
\cite{Tibra2020,Balasubramanian2020,Bhattacharyya2021,Arpan2021,Parker2019,Dymarsky2020,
Caputa2022},
and diffusive transport \cite{Bertini2021}.

The notion of chaos in the classical regime is usually defined by the so-called butterfly effect, 
namely the exponential separation of inifitesimally nearby trajectories 
for initial perturbations \cite{Cvitanovic2005,Schuster2006}.
However, as the concept of trajectory is ill-defined in quantum mechanics, the definition
of quantum chaos remains an open question.
Therefore, to probe the signatures of chaos in quantum many-body systems becomes a central task
in the studies of quantum many-body chaos. 
To date, many complementary detectors of quantum chaos 
and the limits of their usefulness have been widely investigated 
in literature \cite{Chenu2019,BGS1984,Stockmann1999,Haake2010,Zyczkowski1990,
Emerson2002,Bhattacharyya2021,
Arpan2021,Parker2019,Rozenbaum2017,Mata2018,ChenX2018,Kos2018,Bertini2018,
Gietka2019,XuT2020,ZanC2021,Zonnios2022,Alonso2022,Joshi2022}.
Important model systems in this context are billiards \cite{Stockmann1999,Lozej2022}.
Another task, which has recently also drawn great interest, is to unveil different factors 
that affect the chaotic properties of quantum many-body systems.
While the impacts of the strength of disorder and the choice of initial states on the development
of quantum chaos in various many-body systems have been 
extensively explored \cite{Nandkishore2015,Abanin2019,Turner2018,Sinha2020,Turner2021,
Shem2021,Serbyn2021},  more works are required in order to get deeper insights into the universal
aspects of quantum many-body chaos.

In the present work, we analyze the emergence of chaos in the extended Dicke model.
{There are several different versions of the extended Dicke model
\cite{Kloc2017,Rodriguez2018,Guerra2020,Romero2022}.
Here, we focus on the one that has been discussed in Ref.~\cite{Rodriguez2018}. 
Different from the original Dicke model \cite{Dicke1954}, 
which consists of an ensemble of noninteracting two-level atoms interacts with a single bosonic mode,
the atoms in our considered extended Dicke model are permitted to interact.}
This allows us to analyze the effects of the atomic interaction on the degree of chaos of the model.
Previously, the role of the atom-field coupling in the Dicke model 
for the emergence of quantum chaos has been investigated
\cite{Emary2003,QianW2020,Magnani2015,Magnani2016},
while here we explore how this transition is affected by additional atomic interaction.
By performing a detailed analysis of the energy spectral statistics and the structure of eigenstates,
we systematically study both the chaotic signatures of the extended Dicke model
and examine the effect of the atomic interaction on the chaotic features in the model. 
We show how the atomic interaction affects the spectral statistics and 
the structure of eigenstates, respectively.

The article is structured as follows. 
The model is introduced in Sec.~\ref{Secd}.
The influences of the atomic interaction on the energy spectral statistics
are discussed in Sec.~\ref{Third}.
The detailed investigation of the consequence of the atomic interaction 
on the structure of eigenstates is presented in Sec.~\ref{Fourth}. 
Finally, we conclude in Sec.~\ref{Fifth} with a brief summarize of our results and outlook.

 \section{Extended Dicke model} \label{Secd}

 {As an extension of the original Dicke model \cite{Dicke1954,Emary2003,QianW2020}, 
 the extended Dicke model studied here consists of $N$ mutual interacting
 two-level atoms with energy gap $\omega_0$ coupled to 
 a single cavity mode with frequency $\omega$.}
 By employing the collective spin operators 
 $J_{x,y,z}=\sum_{i=1}^N\hat{\sigma}_{x,y,z}^{(i)}$ ($\hat{\sigma}_{x,y,z}$ are the Pauli matrices),
 the Hamiltonian of the extended Dicke model can be written as 
 (hereafter we set $\hbar=1$) \cite{Robles2015,Rodriguez2018}
 \be \label{EDMH}
   H=\omega a^\dag a+\omega_0 J_z+\frac{2\lambda}{\sqrt{N}}J_x(a+a^\dag)
   +\frac{\kappa}{N}J_z^2,
 \ee
 where $a (a^\dag)$ denotes the bosonic annihilation (creation) operator, $\lambda$ 
 is the coupling stength between atom and field, 
 and $\kappa$ represent the strength of the atomic interaction.
 
 The conservation of total spin operator $\mathbf{J}^2=J_x^2+J_y^2+J_z^2$ for the Hamiltonian
 (\ref{EDMH}) leads to the Hamiltonian matrix being block diagonal 
 in $\mathbf{J}^2$ representation.
 In this work, we will focus on the maximum spin sector $j=N/2$, which involves the  
 experimental realizations and includes the ground state.
 Moreover, the commutation between Hamiltonian (\ref{EDMH}) and the parity operator
 $\Pi=e^{i\pi(j+J_z+a^\dag a)}$ enables us to further separate the Hamiltonian matrix into
 even- and odd-parity subspaces. 
 Here, we will restrict our study to the even-parity subspace.  
 
 To numerically diagonalize the Hamiltonian (\ref{EDMH}), we work in the usual Fock-Dicke basis
 $\{|n,m\rangle\}=\{|n\rangle\otimes|j,m\rangle\}$. 
 Here, $|n\rangle$ are the Fock states of bosonic mode with
 $n=0,1,2,\ldots,\infty$ and $|j,m\rangle$ represent the so-called Dicke states with $m=-j,-j+1,\ldots,j$.
 Then, the elements of the Hamiltonian matrix are given by  
 \begin{align}
   \langle n',m'|H|n,m\rangle=
   &(n\omega+m\omega_0)\delta_{n',n}\delta_{m',m}+\frac{\kappa}{N}m^2\delta_{m',m} 
   +\frac{\lambda}{\sqrt{N}}\left[\sqrt{n}\delta_{n',n-1}+\sqrt{n+1}\delta_{n',n+1}\right] \notag \\
   &\times
   \left[\sqrt{j(j+1)-m(m+1)}\delta_{m',m+1}+\sqrt{j(j+1)-m(m-1)}\delta_{m',m-1}\right].
 \end{align}
 We remark that the value of $n$ is unbounded from above, the actual dimension of the 
 Hilbert space is infinite, regardless of the value of $j$.
 In practice, we need to cut off the bosonic number states at a larger but finite value $\mathcal{N}_c$.
 {Moreover, the dependence of the chaoticity in the Dicke model on the energy
 \cite{Rodriguez2018,Pavel2011,Chavez2019} further implies that
 it is also necessary to cut off the energy in order to get the finite number of considered states.
 In our numerical simulations, we set $\mathcal{N}_c=320$ and restrict 
 our analysis on the eigenstates with energies $E/N\in[0.4,4]$, 
 the convergence of our results has been carefully examined.
 For our selected energy interval, we have checked that our main results are still hold
 for other choices of $\mathcal{N}_c$, as long as $N\geq16$ and the convergence of Fock-Dicke
 basis is fulfilled.}
 
 The extended Dicke model exhibits both ground state and excited state 
 quantum phase transitions and displays a transition from integrable to chaotic 
 behavior with increasing the system energy, like in the Dicke model. 
 The features of these transitions have been extensively investigated 
 in the semiclassical regime \cite{Rodriguez2018}.
 It is worth to mention that several possible experimental realizations of the extended Dicke model
 have been pointed out in Refs.~\cite{Lara2011,Robles2015,Rodriguez2018}

 \begin{figure}
    \includegraphics[width=\columnwidth]{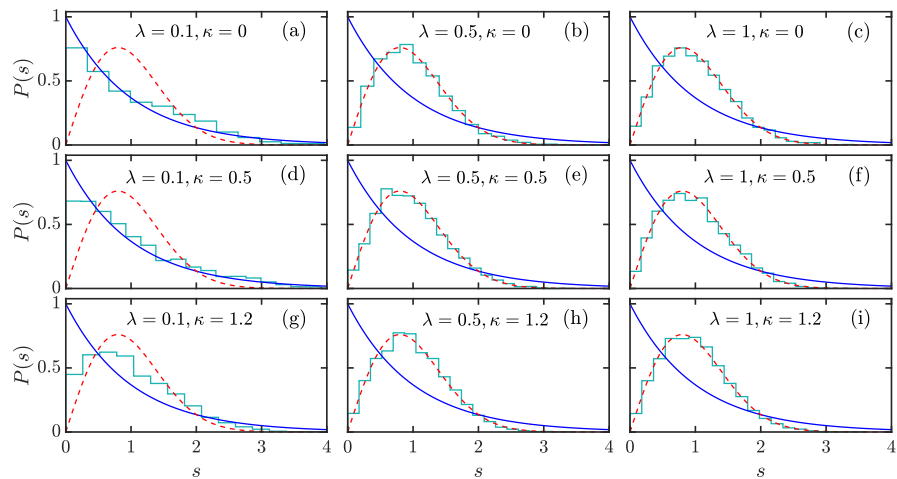}
  \caption{Level spacing distribution of the extended Dicke model for several combinations
  of $\lambda$ and $\kappa$. 
  {The considered energy levels are the ones that have energies $E/N\in[0.4,4]$.}
  The total atom number is $N=2j=32$ and the cut off in bosinic
  Hilbert space is $\mathcal{N}_c=320$. 
  The Poissonian (blue solid lines) and Wigner-Dyson distributions (red dashed lines) are, 
  respectively, plotted in each panel for comparison. 
  Other parameters are: $\omega=\omega_0=1$. All quantities are dimensionless.}
  \label{Psdis}
 \end{figure}

 \section{Energy spectrum statistics} \label{Third}
 
 In this section we will explore the transition from integrable regime to chaos in the extended 
 Dicke model by analyzing its energy level spacing distribution. 
 {In our study, we focus on the energy levels with energies change from $E/N=0.4$ to $E/N=4$.}
 We will compare our results to the level distributions of 
 fully integrable and chaotic cases, respectively.
 We are aiming to characterize the quantum signatures of chaos in the model and 
 unveil the impact of the atomic interaction on its
 chaotic properties.
 
 \subsection{Level spacing statistics}
 
 As the most frequently used probe of quantum chaos, the distribution $P(s)$ of the 
 spacings $s$ of the consecutive unfolded energy levels 
 quantifies the degree of correlations between levels. 
 For integrable systems, where the energy levels are allowed to cross, 
 the distribution $P(s)$ is given by the Poissonian distribution \cite{Berry1977},
 \be
     P_P(s)=\exp(-s).
 \ee
 On the other hand, the energy levels in chaotic systems exhibit level replusion and 
 the distribution $P(s)$ follows the Wigner-Dyson distribution
 \cite{BGS1984,Stockmann1999,Haake2010}.  
 For the systems with symmetric and real Hamiltonian matrices, as in the extended Dicke model,
 the Wigner-Dyson distribution has the following expression
 \be
    P_{WD}(s)=\frac{\pi s}{2}\exp\left(-\frac{\pi}{4}s^2\right).
 \ee 
 
 In Fig.~\ref{Psdis}, we show $P(s)$ of the extended Dicke model 
 with $\omega=\omega_0=1$ and $j=N/2=16$ for different values of $\lambda$ and $\kappa$.
 Here, the level spacings $s$ are obtained from the unfolded eigenlevels 
 $\tilde{E}_\mu=E_\mu/\Delta E$ with $1/\Delta E$ being the local density of states. 
 One can clearly see that $P(s)$ undergoes a transition to Wigner-Dyson distribution as 
 $\lambda$ increases, regardless of the value of $\kappa$. 
 The case $\kappa=0$ is the original Dicke model as in our previous work \cite{QianW2020}. 
 However, we find that, with increasing $\kappa$, 
 the Poissonian distribution at $\lambda=0.1,\kappa=0$ in Fig.~\ref{Psdis}(a) turns
 into an intermediate case, as observed in Fig.~\ref{Psdis}(g). 
 This suggests that the degree of chaos in the extended Dicke model
 can be tuned by the atomic interaction.
 
 To quantitatively characterize the effect of the atomic interaction 
 on the degree of chaos, we consider 
 the proximity of $P(s)$ to Wigner-Dyson or to Poissonian distributions.
 There are several ways to measure the distance between two distributions, here, the difference between
 $P(s)$ and $P_{WD}(s) [P_P(s)]$ is quantified by the chaos indicators $\eta$ and $\beta$. 
 The indicator $\eta$ is defined as \cite{Jacquod1997,Emary2003,March2018}
 \be
   \eta=\left|\frac{\int_0^{s_0}\left[P(s)-P_{WD}(s)\right]ds}
   {\int_0^{s_0}\left[P_P(s)-P_{WD}(s)\right]ds}\right|,
 \ee
 where $s_0=0.4729\ldots$ is the first intersection point of $P_{WD}(s)$ and $P_P(s)$.
 For $P(s)=P_{WD}$, we have $\eta=0$, while $P(s)=P_P(s)$ leads to $\eta=1$.
 The indicator $\beta$ is the level repulsion exponent and can be obtained by fitting $P(s)$ to
 the Brody distribution \cite{Brody1981}
 \be
     P_B(s)=b_\beta(\beta+1)s^\beta\exp[-b_\beta s^{\beta+1}],
 \ee
 where the factor $b_\beta$ is given by
 \be
    b_\beta=\left[\Gamma\left(\frac{\beta+2}{\beta+1}\right)\right]^{\beta+1},
 \ee
 with $\Gamma(x)$ being the gamma function.
 The value of $\beta$ varies in the interval $\beta\in[0,1]$. 
 When $\beta=0$, it means the level spacing distribution $P(s)$ is Poisson.
 On the other hand, for chaotic systems, we would expect $P(s)=P_{WD}(s)$ and  therefore
 $\beta=1$.

  \begin{figure}
    \includegraphics[width=\columnwidth]{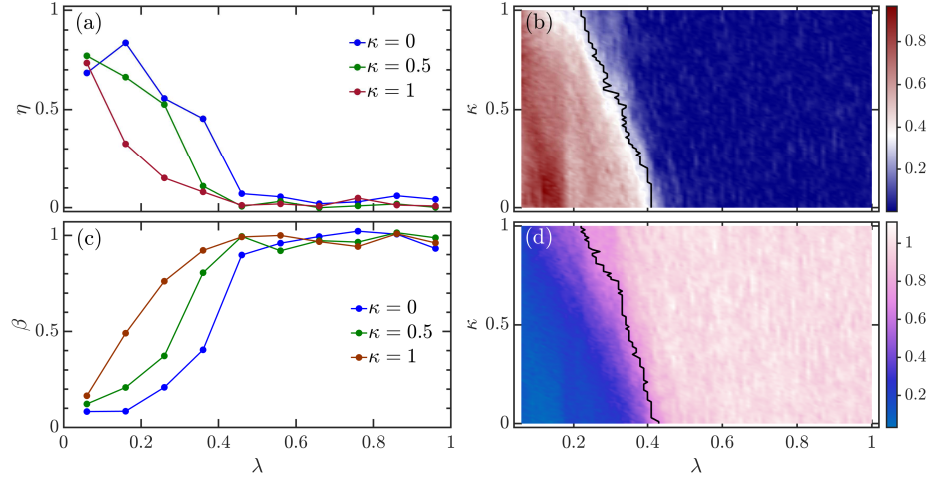}
  \caption{(a) Chaos indicator $\eta$ as a function of $\lambda$ for several values of $\kappa$.
  (b) $\eta$ as a function of $\kappa$ and $\lambda$.
  (c) Chaos indicator $\beta$ as a function of $\lambda$ for different values of $\kappa$.
  (d) $\beta$ as a function of $\kappa$ and $\lambda$.
  {These results are obtained from the energy levels with energies $E/N\in[0.4,4]$.}
  In panels (b) and (d), the boundary of the chaotic region are marked by the black curves, which are
  defined as $\eta\leq0.3$ and $\beta\geq0.7$, respectively. 
  Other parameters are: $\omega=\omega_0=1$ and $N=2j=32$.
  All quantities are dimensionless.}
  \label{Betaeta}
 \end{figure}

 In Fig.~\ref{Betaeta}(a), we plot $\eta$ as a function of $\lambda$ for various values $\kappa$.
 It can be seen that irrespective of the strength of the atomic interaction, 
 the extended Dicke model undergoes a transition from integrability to chaos as the coupling strength 
 $\lambda$ increases.
 However, as the strength of the atomic interaction increases, the onset of chaos
 happens for smaller values of coupling strength $\lambda$.
 This is more evident from Fig.~\ref{Betaeta}(b), where we show how $\eta$ evolves as a function 
 $\kappa$ and $\lambda$.
 Clearly, the width of the region with larger values of $\eta$ decreases with increasing $\kappa$,
 suggesting that the location of the crossover to quantum chaos moves towards the lower values of 
 $\lambda$ with increasing $\kappa$.
 The statement above is further confirmed by the boundary of chaotic region, which  
 plots as the black curve in Fig.~\ref{Betaeta}(b).
 Here, we determine the boundary of chaos by the condition 
 $\eta\leq\eta_d=0.3$ \cite{Jacquod1997}.
 We set the threshold $\eta_d=0.3$ as it implies that the model has already departed from 
 the integrablity and is tending to the chaotic regime.  
 
 Fig.~\ref{Betaeta}(c) shows $\beta$ as a function of $\lambda$ with increasing $\kappa$.
 As observed in Fig.~\ref{Betaeta}(a), while the chaotic behavior at higher values $\lambda$ is
 independent of $\kappa$, the coupling strength $\lambda$ that needs to be for 
 the transition to chaos decreases with increasing $\kappa$.
 Fig.~\ref{Betaeta}(d) illustrates $\beta$ as a function of $\kappa$ and $\lambda$.
 One can see that with increasing $\kappa$ the region with $\beta\approx1$ extends to smaller
 values of $\lambda$. 
 By identifying the chaotic region as $\beta\geq0.7$, we show that the boundary of chaos strongly 
 depends on the strength of the atomic interaction, see the black curve in Fig.~\ref{Betaeta}(d).   
 Notice that the boundary extracted from $\eta$ behaves in a similar way to the one exteracted
 from $\beta$. 
 
 \begin{figure}
    \includegraphics[width=\columnwidth]{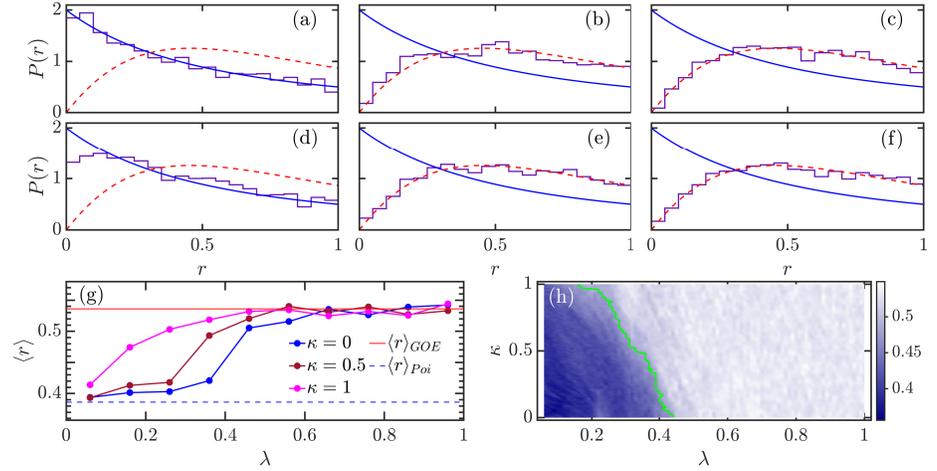}
  \caption{Level spacing ratio distributions $P(r)$ for various combinations of $\kappa$ and $\lambda$:
  (a) $\kappa=0, \lambda=0$, (b) $\kappa=0, \lambda=0.5$, (c) $\kappa=0, \lambda=1$,
  (d) $\kappa=0.7, \lambda=0$, (e) $\kappa=0.7, \lambda=0.5$ and (f) $\kappa=0.7, \lambda=1$. 
  In each panel, the Poisson distribution $P_{Poi}(r)$ is plotted as blue solid curve, 
  while the red dashed line denotes $P_{GOE}(r)$.
  (g) Averaged level spacing ratio $\la r\ra$ as a function of $\lambda$ for several values of $\kappa$. 
  (h) $\langle r\rangle$ as a function of $\kappa$ and $\lambda$.  
  The green line indicates the chaotic boundary, which is determined by $\langle r\rangle\geq0.48$.
  {The energy levels used in our numerical calculation have energies $E/N\in[0.4,4]$.} 
  Other parameters: $\omega=\omega_0=1$ and $N=2j=40$.
  All quantities are dimensionless.}
  \label{PrGap}
 \end{figure}
 
 \subsection{Level spacing ratio}
 
 The study of level spacing distribution requires the so-called unfolding precedure \cite{Guhr1998}. 
 It proceeds by rescaling the original eigenlevels to ensure that the local density of states 
 of the resulting spectrum is $1$. 
 It is usually a non-trivial task, in particular, for quantum many-body systems.
 To circumvent this disadvantage, one can resort to another chaotic probe based on 
 the ratio of adjacent level spacings \cite{Oganesyan2007}, which is free from unfolding procedure.
 
 For a given set of level spacing $\{s_\mu=E_{\mu+1}-E_\mu\}$, the ratio of adjacent level 
 spacings is defined as \cite{Oganesyan2007,Atas2013}
 \be
   r_\mu=\mathrm{min}\left(\delta_\mu,\frac{1}{\delta_\mu}\right),
 \ee
 where $\delta_\mu=s_{\mu+1}/s_\mu$ is the ratio between two adjacent level spacing.
 {Obviously, $r_\mu$ is defined in the interval $r_\mu\in[0,1]$.}
 The distribution of $r_\mu$ for both integrable and chaotic systems has been analytically
 investigated \cite{Atas2013,Atas2013a,Giraud2022}.
 It has been known that for the chaotic systems with Hamitonian from the Gaussian orthogonal
 ensemble (GOE), the level spacing ratio distribution is given by
 \be
    P_{GOE}(r)=\frac{1}{Z_1}\frac{2(r+r^2)}{(1+r+r^2)^{5/2}},
 \ee 
 where $Z_1=8/27$ is the normalization constant.
 On the other hand, as the eigenlevels in the integrable systems are uncorrelated 
 (independent Poisson levels), one can simply find the ratio distribution is
 \be
   P_{Poi}(r)=\frac{2}{(r+1)^2}.
 \ee
 Due to $r_\mu\in[0,1]$, the ratio distribution $P_{GOE/Poi}(r)$ vanishes 
 outside the range $[0,1]$.
 
 Figs.~\ref{PrGap}(a)-\ref{PrGap}(f) show how the level spacing ratio distribution $P(r)$
 evolves for different combinations of atomic interaction strength $\kappa$ and  
 coupling strength $\lambda$.
 Similarly to what we observe for the level spacing distribution $P(s)$ in Fig.~\ref{Psdis}, 
 the spacing ratio distribution $P(r)$ tends to $P_{GOE}$ with 
 increasing coupling strength $\lambda$, independent of $\kappa$ value.
 However, as evident from Figs.~\ref{PrGap}(a) and \ref{PrGap}(d), increasing $\kappa$ leads to
 the enhancement in degree of chaos of the model at smaller values of $\lambda$.
 Therefore, as mentioned above, the atomic interaction can be used 
 to tune the level of chaoticity in the model.
 By switching on the interatomic interaction $\kappa>0$, the regularity-to-chaos transition of
 the original Dicke model \cite{Magnani2016,QianW2020} is amplified.
 
 \begin{figure}
    \includegraphics[width=\columnwidth]{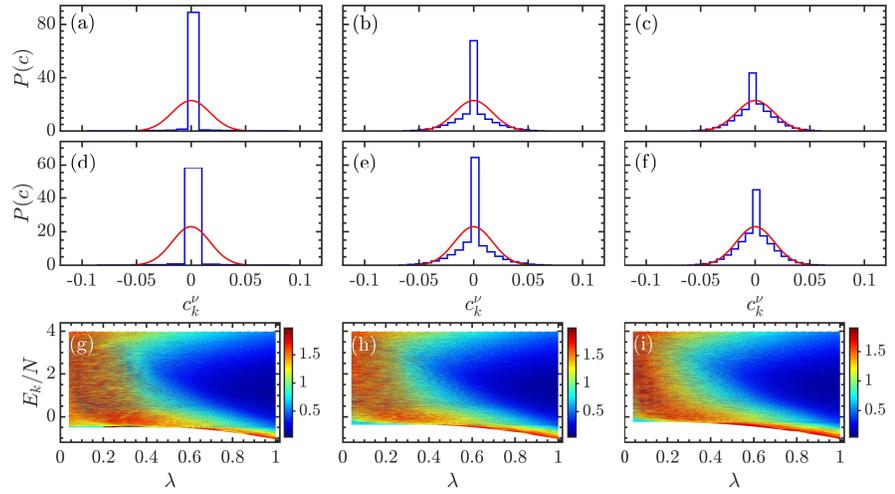}
  \caption{Panels (a)-(f): Histogramas of the coefficients $c_k^\nu$ of eigenstates with energy
  $E_k/N\in[1.75,2.25]$, in the Fock-Dicke basis.
  Each panel corresponds to different $(\kappa,\lambda)$ combinations:  
  (a) $\kappa=0,\lambda=0$, (b) $\kappa=0,\lambda=0.5$, 
  (c) $\kappa=0,\lambda=1$, (d) $\kappa=0.7,\lambda=0$, 
  (e) $\kappa=0.7,\lambda=0.5$, and (f) $\kappa=0.7,\lambda=1$.
  The red solid line in each panel denotes the Gaussian distribution in Eq.~(\ref{Gsdf}).
  Panels (g)-(i): Kullback-Leibler divergence $D_{KL}$ [cf.~Eq.~(\ref{DKLD})]
  as a function of rescaled energy $E_k/N$ and $\lambda$
  for $\kappa=0$ (g), $\kappa=0.5$ (h), and $\kappa=1$ (i).
  Other parameters: $\omega=\omega_0=1$ and $N=2j=40$.
  All quantities are dimensionless.}
  \label{DisEgs}
 \end{figure}

 A more stringent analysis of the effect of the atomic interaction is made with the 
 average level spacing ratio, defined as
 \be
   \langle r\rangle=\int_0^1 rP(r)dr.
 \ee
 It takes the value $\langle r\rangle_{Poi}=2\ln 2-1\approx0.386$ for integrable systems 
 with $P(r)=P_{Poi}(r)$, while for chaotic systems with $P(r)=P_{GOE}(r)$, one has
 $\langle r\rangle_{GOE}=4-2\sqrt{3}\approx0.536$.
 Hence, $\langle r\rangle$ acts as a detector to diagnose 
 whether the studied system is in the integrable or chaotic regime and has been
 widely used to track the crossover from integrability to chaos.
 
 Fig.~\ref{PrGap}(g) demonstrates $\langle r\rangle$ as a function of $\lambda$ for three different 
 values $\kappa$.
 We see that, regardless of the value of $\kappa$, the transition from integrability
 to chaos is well captured by the behavior of $\langle r\rangle$, which varies from 
 $\langle r\rangle_{Poi}$ to $\langle r\rangle_{GOE}$ with increasing $\lambda$.
 We further observe that the chaotic phase is robust with respect to variation of $\kappa$,
 but the integrable phase exhibits a strong dependence on $\kappa$.
 For the integrable phase with smaller values of $\lambda$, 
 we find that increasing $\kappa$ gives rise to an increase of $\langle r\rangle$. 
 As a consequence, the location of transition to chaos can be varied
 by the atomic intraction.
 This effect is more clearly observed in Fig.~\ref{PrGap}(h) where the evolves of $\langle r\rangle$
 as a function of $\kappa$ and $\lambda$ has been illustrated.
 Again, we define the boundary of chaotic region by the condition 
 $\langle r\rangle\geq\langle r\rangle_c=0.48$, meaning that the region with value 
 $\langle r\rangle\geq 0.48$ is considered as chaos.
 We have checked that our main result still holds for other choices of $\langle r\rangle_c$, 
 as long as $\langle r\rangle_c\in(0.45,0.5)$.
 The green line in Fig.~\ref{PrGap}(h) denotes the obtained boundary of chaos.
 As expected, the behavior of the boundary line confirms the
 extension of the chaotic region to lower values of $\lambda$ as $\kappa$ increases.

 \section{Structure of eigenstates} \label{Fourth}
 
 The onset of chaos also bears a remarkable change in the structure of eigenstates.
 In this section we explore the impact of atomic interaction on the transition to chaos by investigating
 the variation in the structure of eigenstates. 
 
 It is known that the eigenstates of chaotic systems are uncorrelated and are
 well described by random matrix theory (RMT)
 \cite{Borgonovi2016,Berry1977a,Porter1956,Mehta2004}.
 For the model studied in this work, one can expect that in the chaotic phase
 the eigenstates of the model will have the same structure as those of random GOE matrices.
 The GOE eigenstates are fully delocalized random vectors with real components consist of 
 independent Gaussian random numbers.
 Hence, the deviation of eigenvector structure from Gaussian behavior 
 is an alternative benchmark to certify quantum chaos  
 \cite{Izrailev1990,HaakeZ1990,Zyczkowski1990,Leboeuf1990a,QianWg2021}. 
 
  \begin{figure}
    \includegraphics[width=\columnwidth]{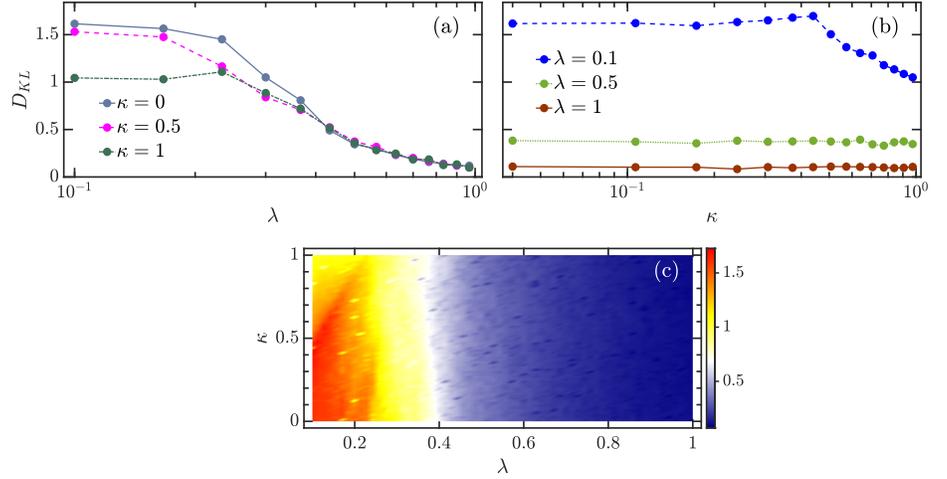}
  \caption{(a) KL divergence $D_{KL}$ (\ref{DKLD}) as a function of $\lambda$ 
  for several $\kappa$ values.
  (b) $D_{KL}$ as a function of $\kappa$ for several coupling strengths $\lambda$.
  (c) Color scaled plot of $D_{KL}$ in the $\kappa-\lambda$ plane. 
  In all panels, the coefficients distribution $P(c)$ is obtained from the eigenstates with energies 
  $E_k/N\in[1.75,2.25]$.
  Other parameters: $\omega=\omega_0=1$ and $N=2j=40$.
  All quantities are dimensionless.}
  \label{Devpdf}
 \end{figure}

The analysis of the structure of eigenstates requires expansion of the eigenstates in a chosen basis.
The choice of basis is usually decided by the physical problem and the system under consideration.
Here, we use the Fock-Dicke basis, $\{|n,m\rangle\}$, 
which are the eigenstates of $a^\dag a$ and $J_z$, as mentioned in Sec.~\ref{Secd}.
The decomposition of the $k$th eigenstate, $|k\rangle$, of the Hamiltonian (\ref{EDMH})
in the selected basis is given by
\be \label{ExpEgk}
  |k\rangle=\sum_{\nu=1}^\mathcal{D}c_k^\nu|\nu\rangle,
\ee
where $|\nu\rangle=|n,m\rangle$, $\mathcal{D}$ is the dimension of the Hilbert space, and
$c_{k}^\nu=\langle\nu|k\rangle$ are the $k$th eigenstate components 
in the basis $\{|\nu\rangle\}$, 
satisfying the normalization condition {$\sum_\nu|c_{k}^{\nu}|^2=1$.}
The characterizations of the eigenstate structure are provided by the statistical properties of
eigenstate coefficients $\{c_{k}^{\nu}\}$. 

To analyze the fingerprint of chaos in the properties of the eigenstates of $H$ in Eq.~(\ref{EDMH}), 
as well as to examine whether the atomic interaction has effect on the eigenstate structure, 
we explore the coefficients distribution in comparison with the corresponding GOE result.
As mentioned above, the eigenstates for the chaotic systems 
with real and symmetric $\mathcal{D}$-dimensional 
Hamiltonian matrices are consistent with the GOE eigenstates.
The components of a GOE eigenstate, $\{c_\nu\}$, are random numbers that are uniformly 
distributed on a unit sphere with dimension $\mathcal{D}-1$. 
In the limit $\mathcal{D}\gg1$, the dependence between components vanishes and the 
distribution of components can be well described by a Gaussian distribution with 
zero mean and variance $1/\mathcal{D}$ 
\cite{Brody1981,Herrera2016,Backer2019,Nakerst2022},
\be \label{Gsdf}
  P_{GOE}(c)=\sqrt{\frac{\mathcal{D}}{2\pi}}e^{-\mathcal{D}c^2/2}.
\ee
In the chaotic systems, it has been known that the coefficients of 
the mid-spectrum eigenstates are distributed as near-Gaussian distribution
\cite{Brody1981,Guhr1998,Khaymovich2019,Pausch2021,Wouter2018},
while the coefficients distribution for eigenstates of non-chaotic systems and 
the edge eigenstates of chaotic systems is significantly different from 
Gaussian distribution \cite{Wouter2018,Luitz2016,Luitz2020}.
As increasing $\lambda$ leads to the onset of chaos in the model, 
one would expect that the distribution of mid-spectrum eigenstates coefficients 
shoud be turned from non-Gaussian into near-Gaussian.

Figs.~\ref{DisEgs}(a)-\ref{DisEgs}(f) show the evolution of the eigenstates coefficients
distributions, denoted by $P(c)$,
for several $(\kappa,\lambda)$ combinations, in comparsion with 
the Gaussian distribution provided by Eq.~(\ref{Gsdf}).
We see that, regardless of the value of $\kappa$, the eigenstates coefficients distribution
tends to Gaussian as $\lambda$ increases.
The larger the value of $\lambda$ is, the higher the degree of chaos in the model, and, therefore, the
closer the coefficients distribution to Gaussian, as expected.
Another prominent feature, that is also independent of $\kappa$, 
observed in the behaviors of $P(c)$ is the larger peak around $c_k^\nu\sim0$.
Even in the most chaotic regime, $P(c)$ still exhibits a high peak near zero,
as shown in Figs.~\ref{DisEgs}(c) and \ref{DisEgs}(f). 
This excessive number of zero coefficients is mainly due to the mixed feature of the model.
This means the regular and chaotic behavior coexist in the model for the considered parameters.
Detailed analysis of the mixture of regular and chaotic behaviors in extended 
Dicke model is beyond the scope
of the present work, we leave this investigation for a furture work.

Let us now turn to discuss how the atomic interaction $\kappa$ 
affects the eigenstate coefficients distribution. 
As evident from Figs.~\ref{DisEgs}(a)-\ref{DisEgs}(f), the presence of atomic interaction 
has almost no effect on the behavior of eigenstate coefficients distribution, 
but just a reduction of the hight of peak in the regular phase 
[compare Figs.~\ref{DisEgs}(a) to \ref{DisEgs}(d)].
This implies that the impact of the atomic interaction on the structure of eigenstates is not as strong
as on the eigenlevels.

To measure the difference between $P(c)$ and Gaussian distribution (\ref{Gsdf}), 
we consider Kullback-Leibler (KL) divergence \cite{Kullback1951}, 
which is commonly used to measure how close an
observed distribution is to a predicted distribution. 
In our study, $P(c)$ and $P_{GOE}(c)$ are, respectively, identified 
as the observed and predicted distributions.
Then, the KL divergence between them is given by
\be\label{DKLD}
   D_{KL}=\int_{c_{min}}^{c_{max}} P(c)\ln\left[\frac{P(c)}{P_{GOE}(c)}\right]dc,
\ee
where $c_{min}$ and $c_{max}$ denote respectively the minimum 
and maximum values in $\{c_k^\nu\}$.
The KL divergence is non-negative, so that $D_{KL}\geq0$, 
and it vanishes only when $P(c)=P_{GOE}(c)$.
Qualitatively, a larger $D_{KL}$ value indicates a larger difference between $P(c)$ and $P_{GOE}(c)$. 

In Figs.~\ref{DisEgs}(g)-\ref{DisEgs}(i), we plot the KL divergence for the extended Dicke model
as a function of $E_k/N$ and $\lambda$ with $\kappa=\{0,0.5,1\}$.
We see that the behaviors of $D_{KL}$ versus $E_k/N$ and $\lambda$ 
for different $\kappa$ values are very similar. 
As expected, the eigenstates in the non-chaotic phase $(\lambda\lesssim0.45)$ and the eigenstates 
at the spectrum edge in the chaotic phase $(\lambda\gtrsim0.5)$ have larger values of $D_{KL}$,
suggesting that the corresponding coefficients distributions 
strongly deviate from Gaussian, as illustrated in
Figs.~\ref{DisEgs}(a) and \ref{DisEgs}(d) for integrable case.
Here, it is worth pointing out that the infinite dimension of the Hilbert space for the extended Dicke
model leads to its spectrum has only one edge, namely the ground state.
On the other hand, the lower values of $D_{KL}$ for mid-spectrum eigenstates in the chaotic phase
imply that their coefficients distributions are closer to Gaussian, 
as shown in Figs.~\ref{DisEgs}(c) and \ref{DisEgs}(f). 
The similarity between behaviors of $D_{KL}$ observed in the bottom row of Fig.~(\ref{DisEgs})
prompts a more detailed investigation of the impact of the atomic interaction 
on the eigenstate coefficient distribution.

To determine whether the eigenstate coefficients distribution is robust with respect to 
the variation of the atomic interaction $\kappa$, we calculate $D_{KL}$ for the 
mid-spectrum eigenstates with energies $E_k/N\in[1.75,2.25]$.
The final result for various cases is shown in Fig.~\ref{Devpdf}.
The evolution of $D_{KL}$ as a function of $\lambda$ for different
values of $\kappa$ is plotted in Fig.~\ref{Devpdf}(a).
One can see that the behavior of $D_{KL}$ for different $\kappa$ is very similar.
The KL divergence varies slowly for smaller $\lambda$ until $\lambda\approx0.25$, after which
it rapidly decreases to small values as the coupling strength is increased 
beyond $\lambda_c\approx0.5$. 
The change of $D_{KL}$ behavior is a manifestation of the
transition to chaos resulting from the reduction of eigenstates correlation with increasing $\lambda$.
We also observe that the dependence of $D_{KL}$ on $\kappa$ in the integrable phase is very
different from that of chaotic phase. 
The explicit dependence of $D_{KL}$ on $\kappa$ for several values of $\lambda$ is shown in
Fig.~\ref{Devpdf}(b).
As expected from Fig.~\ref{Devpdf}(a), in the chaotic regime with $\lambda\gtrsim0.5$,
$D_{KL}$ is independent of $\kappa$,
whereas the KL divergence strongly depends on the atomic interaction 
for the cases of smaller $\lambda$.  
Overall, the KL divergence decreases with increasing $\kappa$ in the integrable regime.
This indicates that the proximity of the eigenstate coefficients distribution to the Gaussian distribution
can be improved by the atomic interaction. 
Therefore, the atomic interaction can vary the degree of chaos in the extended Dicke model, in
agreement with the results obtained from eigenvalue statistics.
An overall evolution of $D_{KL}$ as a function of $\kappa$ and $\lambda$ is depicted in
Fig.~\ref{Devpdf}(c).
We see that the chaotic region remains almost unchanged as $\kappa$ increases, 
in contrast to the extension behavior revealed by the eigenvalue-based detectors of quantum chaos
[see Figs.~\ref{Betaeta}(b), \ref{Betaeta}(d), and \ref{PrGap}(h)].
This suggests that the level repulsion is more sensitive to the effect of atomic interaction than the 
structure of eigenstates.

 \section{Conclusions} \label{Fifth}
 
 In this article we have performed a detailed analysis of quantum chaotic characters  
 of the extended Dicke model through the statistical properties of eigenvalues and eigenstates.
 The presence of interaction between atoms in the model further allows us to
 explore the dependence of chaotic properties of the model on the atomic interaction.
 It has been shown that the integrability-to-chaos transition of the original Dicke model
 as a function of the atom-field coupling is amplified by the interatomic interaction.
 
 We have demonstrated that as the model moves from regular phase to chaotic phase,
 both the level spacing and level spacing ratio distributions undergo a crossover from
 Poisson to Wigner distribution, regardless of the strength of atomic interaction.
 However, the presence of atomic interaction can lead to a notable deviation of 
 level spacing and spacing ratio distributions from Poisson distribution.
 To quantify this deviation and to measure the degree of chaos in the model,
 we consider three different chaos indictors to probe the transition from integrablity to chaos.
 All of these indicators are complementary to each other and are able to capture the
 crossover from Poisson to Wigner for both level spacing and level spacing ratio distributions.
 We have also shown that the behaviors of these indicators as a function of 
 control parameters are very similar.  
 In particular, we found that the degree of chaos of the model can be controlled
 by tuning the strength of atomic interaction. 
 This result highlights the role of interaction in the development of 
 chaos in quantum many-body systems and opens up the possiblity to tune the degree of chaos
 in the extended Dicke model.
 
 Further quantum chaotic signatures showing how the atomic interaction 
 affects the degree of chaos in the
 extended Dicke model are unveiled in the structure of the eigenstates.
 To analyze the eigenstate structure, we expand each eigenstate in the Fock-Dicke basis and
 focus on the expansion coefficients distribution for mid-spectrum eigenstates.  
 For fully chaotic systems with Hamitonian from GOE, such distribution is 
 well described by Gaussian distribution.
 We have shown that the transition to chaos can be detected by the deviation 
 of coefficients distribution from Gaussian distribution.
 However, we note that even within the chaotic phase, the coefficients distribution is still different
 from Gaussian, indicating the existence of correlations between them.
 By using the KL divergence to measure the distance between the coefficients distribution and
 Gaussian distribution, we have illustrated that the onset of chaos corresponds to the rapid decrease
 in the behavior of KL divergence as a function of coupling strength.
 Although the atomic interaction leads to the decrease of KL divergence in the regular phase, 
 the transition to chaos revealed by KL divergence is almost independent of atomic interaction.
 This is different from the results obtained by the eigenvalue-based chaos indicators and
 implies that unlike the eigenlevels, the eigenstate structure is robust 
 with respect to the change of atomic interaction.
 
 A nutural extension of the present work is to investigate the 
 dynamical role played by the atomic interaction in the development of chaos.
 It would also be interesting to analyze the effect of atomic interaction on the level of chaoticity
 through the long-range spectral correlations, 
 which can be detected by the spectral form factor \cite{Haake2010}.
 In addition, understanding the emergence of chaos and the impact of atomic interaction from the
 dynamics of the classical counterpart of the model would be another interesting topic.
 {Very recently, the critical phenomena in the extended Dicke model have been thoroughly
 analyzed in this direction \cite{Romero2022}.}
 Finally, we would like to mention that a direct demonstration of level spacing distribution 
 in an ultracold-atom system has been realized in a recent experiment \cite{Frisch2014}. 
 Hence, we expect that the spectral statistics of our studied extended Dicke model can be verified
 by the state-of-the-art experimental platforms.   
         

\funding{This research was funded by the Slovenian Research Agency (ARRS) under the grant 
number J1-9112. 
Q.~W. acknowledges support from the National Science Foundation of China under grant number 11805165,
Zhejiang Provincial Nature Science Foundation under grant number LY20A050001.}

\institutionalreview{Not applicable.}

\informedconsent{Not applicable.}

\dataavailability{Not applicable.} 

\conflictsofinterest{The authors declare no conflict of interest.}

\end{paracol}
 
\reftitle{References}   
 
\externalbibliography{yes}
\bibliography{QCEDM}

\begin{thebibliography}{999}

\bibitem[Altland and Haake(2012)]{Altland2012}
Altland, A.; Haake, F.
\newblock Quantum Chaos and Effective Thermalization.
\newblock {\em Phys. Rev. Lett.} {\bf 2012}, {\em 108},~073601.
\newblock
  doi:{\changeurlcolor{black}\href{https://doi.org/10.1103/PhysRevLett.108.073601}{\detokenize{10.1103/PhysRevLett.108.073601}}}.

\bibitem[D'Alessio \em{et~al.}(2016)D'Alessio, Kafri, Polkovnikov, and
  Rigol]{LucaD2016}
D'Alessio, L.; Kafri, Y.; Polkovnikov, A.; Rigol, M.
\newblock From quantum chaos and eigenstate thermalization to statistical
  mechanics and thermodynamics.
\newblock {\em Adv. Phys.} {\bf 2016}, {\em 65},~239--362.
\newblock
  doi:{\changeurlcolor{black}\href{https://doi.org/10.1080/00018732.2016.1198134}{\detokenize{10.1080/00018732.2016.1198134}}}.

\bibitem[Borgonovi \em{et~al.}(2016)Borgonovi, Izrailev, Santos, and
  Zelevinsky]{Borgonovi2016}
Borgonovi, F.; Izrailev, F.; Santos, L.; Zelevinsky, V.
\newblock Quantum chaos and thermalization in isolated systems of interacting
  particles.
\newblock {\em Phys. Rep.} {\bf 2016}, {\em 626},~1--58.
\newblock
  doi:{\changeurlcolor{black}\href{https://doi.org/https://doi.org/10.1016/j.physrep.2016.02.005}{\detokenize{https://doi.org/10.1016/j.physrep.2016.02.005}}}.

\bibitem[Nandkishore and Huse(2015)]{Nandkishore2015}
Nandkishore, R.; Huse, D.A.
\newblock Many-Body Localization and Thermalization in Quantum Statistical
  Mechanics.
\newblock {\em Ann. Rev. Condens. Matter Phys.} {\bf 2015}, {\em 6},~15--38.
\newblock
  doi:{\changeurlcolor{black}\href{https://doi.org/10.1146/annurev-conmatphys-031214-014726}{\detokenize{10.1146/annurev-conmatphys-031214-014726}}}.

\bibitem[Deutsch(2018)]{Deutsch2018}
Deutsch, J.M.
\newblock Eigenstate thermalization hypothesis.
\newblock {\em Rep. Prog. Phys.} {\bf 2018}, {\em 81},~082001.
\newblock
  doi:{\changeurlcolor{black}\href{https://doi.org/10.1088/1361-6633/aac9f1}{\detokenize{10.1088/1361-6633/aac9f1}}}.

\bibitem[Chan \em{et~al.}(2018)Chan, De~Luca, and Chalker]{ChanA2018}
Chan, A.; De~Luca, A.; Chalker, J.T.
\newblock Solution of a Minimal Model for Many-Body Quantum Chaos.
\newblock {\em Phys. Rev. X} {\bf 2018}, {\em 8},~041019.
\newblock
  doi:{\changeurlcolor{black}\href{https://doi.org/10.1103/PhysRevX.8.041019}{\detokenize{10.1103/PhysRevX.8.041019}}}.

\bibitem[Garcia-March \em{et~al.}(2018)Garcia-March, van Frank, Bonneau,
  Schmiedmayer, Lewenstein, and Santos]{March2018}
Garcia-March, M.A.; van Frank, S.; Bonneau, M.; Schmiedmayer, J.; Lewenstein,
  M.; Santos, L.F.
\newblock Relaxation, chaos, and thermalization in a three-mode model of a
  Bose{\textendash}Einstein condensate.
\newblock {\em New Journal of Physics} {\bf 2018}, {\em 20},~113039.
\newblock
  doi:{\changeurlcolor{black}\href{https://doi.org/10.1088/1367-2630/aaed68}{\detokenize{10.1088/1367-2630/aaed68}}}.

\bibitem[Friedman \em{et~al.}(2019)Friedman, Chan, De~Luca, and
  Chalker]{Friedman2019}
Friedman, A.J.; Chan, A.; De~Luca, A.; Chalker, J.T.
\newblock Spectral Statistics and Many-Body Quantum Chaos with Conserved
  Charge.
\newblock {\em Phys. Rev. Lett.} {\bf 2019}, {\em 123},~210603.
\newblock
  doi:{\changeurlcolor{black}\href{https://doi.org/10.1103/PhysRevLett.123.210603}{\detokenize{10.1103/PhysRevLett.123.210603}}}.

\bibitem[Ray \em{et~al.}(2020)Ray, Cohen, and Vardi]{Ray2020}
Ray, S.; Cohen, D.; Vardi, A.
\newblock Chaos-induced breakdown of Bose-Hubbard modeling.
\newblock {\em Phys. Rev. A} {\bf 2020}, {\em 101},~013624.
\newblock
  doi:{\changeurlcolor{black}\href{https://doi.org/10.1103/PhysRevA.101.013624}{\detokenize{10.1103/PhysRevA.101.013624}}}.

\bibitem[Rautenberg and G\"arttner(2020)]{Garttner2020}
Rautenberg, M.; G\"arttner, M.
\newblock Classical and quantum chaos in a three-mode bosonic system.
\newblock {\em Phys. Rev. A} {\bf 2020}, {\em 101},~053604.
\newblock
  doi:{\changeurlcolor{black}\href{https://doi.org/10.1103/PhysRevA.101.053604}{\detokenize{10.1103/PhysRevA.101.053604}}}.

\bibitem[Kobrin \em{et~al.}(2021)Kobrin, Yang, Kahanamoku-Meyer, Olund, Moore,
  Stanford, and Yao]{Kobrin2021}
Kobrin, B.; Yang, Z.; Kahanamoku-Meyer, G.D.; Olund, C.T.; Moore, J.E.;
  Stanford, D.; Yao, N.Y.
\newblock Many-Body Chaos in the Sachdev-Ye-Kitaev Model.
\newblock {\em Phys. Rev. Lett.} {\bf 2021}, {\em 126},~030602.
\newblock
  doi:{\changeurlcolor{black}\href{https://doi.org/10.1103/PhysRevLett.126.030602}{\detokenize{10.1103/PhysRevLett.126.030602}}}.

\bibitem[Fogarty \em{et~al.}(2021)Fogarty, Garc{\'{i}}a-March, Santos, and
  Harshman]{Fogarty2021}
Fogarty, T.; Garc{\'{i}}a-March, M.{\'{A}}.; Santos, L.F.; Harshman, N.L.
\newblock Probing the edge between integrability and quantum chaos in
  interacting few-atom systems.
\newblock {\em {Quantum}} {\bf 2021}, {\em 5},~486.
\newblock
  doi:{\changeurlcolor{black}\href{https://doi.org/10.22331/q-2021-06-29-486}{\detokenize{10.22331/q-2021-06-29-486}}}.

\bibitem[Wittmann~W. \em{et~al.}(2022)Wittmann~W., Castro, Foerster, and
  Santos]{Wittmann2022}
Wittmann~W., K.; Castro, E.R.; Foerster, A.; Santos, L.F.
\newblock Interacting bosons in a triple well: Preface of many-body quantum
  chaos.
\newblock {\em Phys. Rev. E} {\bf 2022}, {\em 105},~034204.
\newblock
  doi:{\changeurlcolor{black}\href{https://doi.org/10.1103/PhysRevE.105.034204}{\detokenize{10.1103/PhysRevE.105.034204}}}.

\bibitem[Maldacena \em{et~al.}(2016)Maldacena, Shenker, and
  Stanford]{Maldacena2016}
Maldacena, J.; Shenker, S.H.; Stanford, D.
\newblock A bound on chaos.
\newblock {\em J. High Energy Phys.} {\bf 2016}, {\em 2016},~106.
\newblock
  doi:{\changeurlcolor{black}\href{https://doi.org/10.1007/JHEP08(2016)106}{\detokenize{10.1007/JHEP08(2016)106}}}.

\bibitem[Stanford(2016)]{Stanford2016}
Stanford, D.
\newblock Many-body chaos at weak coupling.
\newblock {\em J. High Energy Phys.} {\bf 2016}, {\em 2016},~9.
\newblock
  doi:{\changeurlcolor{black}\href{https://doi.org/10.1007/JHEP10(2016)009}{\detokenize{10.1007/JHEP10(2016)009}}}.

\bibitem[Mag{\'a}n(2018)]{Magan2018}
Mag{\'a}n, J.M.
\newblock Black holes, complexity and quantum chaos.
\newblock {\em J. High Energy Phys.} {\bf 2018}, {\em 2018},~43.
\newblock
  doi:{\changeurlcolor{black}\href{https://doi.org/10.1007/JHEP09(2018)043}{\detokenize{10.1007/JHEP09(2018)043}}}.

\bibitem[Jahnke(2019)]{Jahnke2019}
Jahnke, V.
\newblock Recent Developments in the Holographic Description of Quantum Chaos.
\newblock {\em Adv. High Energy Phys.} {\bf 2019}, {\em 2019},~9632708.
\newblock
  doi:{\changeurlcolor{black}\href{https://doi.org/10.1155/2019/9632708}{\detokenize{10.1155/2019/9632708}}}.

\bibitem[Ali \em{et~al.}(2020)Ali, Bhattacharyya, Haque, Kim, Moynihan, and
  Murugan]{Tibra2020}
Ali, T.; Bhattacharyya, A.; Haque, S.S.; Kim, E.H.; Moynihan, N.; Murugan, J.
\newblock Chaos and complexity in quantum mechanics.
\newblock {\em Phys. Rev. D} {\bf 2020}, {\em 101},~026021.
\newblock
  doi:{\changeurlcolor{black}\href{https://doi.org/10.1103/PhysRevD.101.026021}{\detokenize{10.1103/PhysRevD.101.026021}}}.

\bibitem[Rabinovici \em{et~al.}(2021)Rabinovici, Sánchez-Garrido, Shir, and
  Sonner]{Rabinovici2021}
Rabinovici, E.; Sánchez-Garrido, A.; Shir, R.; Sonner, J.
\newblock Operator complexity: a journey to the edge of Krylov space.
\newblock {\em J. High Energy Phys.} {\bf 2021}, {\em 2021},~62.

\bibitem[Schack and Caves(1996)]{Schack1996}
Schack, R.; Caves, C.M.
\newblock Information-theoretic characterization of quantum chaos.
\newblock {\em Phys. Rev. E} {\bf 1996}, {\em 53},~3257--3270.
\newblock
  doi:{\changeurlcolor{black}\href{https://doi.org/10.1103/PhysRevE.53.3257}{\detokenize{10.1103/PhysRevE.53.3257}}}.

\bibitem[Vidmar and Rigol(2017)]{Vidmar2017}
Vidmar, L.; Rigol, M.
\newblock Entanglement Entropy of Eigenstates of Quantum Chaotic Hamiltonians.
\newblock {\em Phys. Rev. Lett.} {\bf 2017}, {\em 119},~220603.
\newblock
  doi:{\changeurlcolor{black}\href{https://doi.org/10.1103/PhysRevLett.119.220603}{\detokenize{10.1103/PhysRevLett.119.220603}}}.

\bibitem[Piga \em{et~al.}(2019)Piga, Lewenstein, and Quach]{Piga2019}
Piga, A.; Lewenstein, M.; Quach, J.Q.
\newblock Quantum chaos and entanglement in ergodic and nonergodic systems.
\newblock {\em Phys. Rev. E} {\bf 2019}, {\em 99},~032213.
\newblock
  doi:{\changeurlcolor{black}\href{https://doi.org/10.1103/PhysRevE.99.032213}{\detokenize{10.1103/PhysRevE.99.032213}}}.

\bibitem[Bertini \em{et~al.}(2019)Bertini, Kos, and Prosen]{Bertini2019}
Bertini, B.; Kos, P.; Prosen, T.c.v.
\newblock Entanglement Spreading in a Minimal Model of Maximal Many-Body
  Quantum Chaos.
\newblock {\em Phys. Rev. X} {\bf 2019}, {\em 9},~021033.
\newblock
  doi:{\changeurlcolor{black}\href{https://doi.org/10.1103/PhysRevX.9.021033}{\detokenize{10.1103/PhysRevX.9.021033}}}.

\bibitem[Lerose and Pappalardi(2020)]{Lerose2020}
Lerose, A.; Pappalardi, S.
\newblock Bridging entanglement dynamics and chaos in semiclassical systems.
\newblock {\em Phys. Rev. A} {\bf 2020}, {\em 102},~032404.
\newblock
  doi:{\changeurlcolor{black}\href{https://doi.org/10.1103/PhysRevA.102.032404}{\detokenize{10.1103/PhysRevA.102.032404}}}.

\bibitem[Lantagne-Hurtubise \em{et~al.}(2020)Lantagne-Hurtubise, Plugge, Can,
  and Franz]{Lantagne2020}
Lantagne-Hurtubise, E.; Plugge, S.; Can, O.; Franz, M.
\newblock Diagnosing quantum chaos in many-body systems using entanglement as a
  resource.
\newblock {\em Phys. Rev. Research} {\bf 2020}, {\em 2},~013254.
\newblock
  doi:{\changeurlcolor{black}\href{https://doi.org/10.1103/PhysRevResearch.2.013254}{\detokenize{10.1103/PhysRevResearch.2.013254}}}.

\bibitem[Anand \em{et~al.}(2021)Anand, Styliaris, Kumari, and
  Zanardi]{Anand2021}
Anand, N.; Styliaris, G.; Kumari, M.; Zanardi, P.
\newblock Quantum coherence as a signature of chaos.
\newblock {\em Phys. Rev. Research} {\bf 2021}, {\em 3},~023214.
\newblock
  doi:{\changeurlcolor{black}\href{https://doi.org/10.1103/PhysRevResearch.3.023214}{\detokenize{10.1103/PhysRevResearch.3.023214}}}.

\bibitem[Hosur \em{et~al.}(2016)Hosur, Qi, Roberts, and Yoshida]{Hosur2016}
Hosur, P.; Qi, X.L.; Roberts, D.A.; Yoshida, B.
\newblock Chaos in quantum channels.
\newblock {\em Journal of High Energy Physics} {\bf 2016}, {\em 2016},~4.

\bibitem[Chenu \em{et~al.}(2019)Chenu, Molina-Vilaplana, and del
  Campo]{Chenu2019}
Chenu, A.; Molina-Vilaplana, J.; del Campo, A.
\newblock Work {S}tatistics, {L}oschmidt {E}cho and {I}nformation {S}crambling
  in {C}haotic {Q}uantum {S}ystems.
\newblock {\em {Quantum}} {\bf 2019}, {\em 3},~127.
\newblock
  doi:{\changeurlcolor{black}\href{https://doi.org/10.22331/q-2019-03-04-127}{\detokenize{10.22331/q-2019-03-04-127}}}.

\bibitem[Prakash and Lakshminarayan(2020)]{Prakash2020}
Prakash, R.; Lakshminarayan, A.
\newblock Scrambling in strongly chaotic weakly coupled bipartite systems:
  Universality beyond the Ehrenfest timescale.
\newblock {\em Phys. Rev. B} {\bf 2020}, {\em 101},~121108.
\newblock
  doi:{\changeurlcolor{black}\href{https://doi.org/10.1103/PhysRevB.101.121108}{\detokenize{10.1103/PhysRevB.101.121108}}}.

\bibitem[Balasubramanian \em{et~al.}(2020)Balasubramanian, DeCross, Kar, and
  Parrikar]{Balasubramanian2020}
Balasubramanian, V.; DeCross, M.; Kar, A.; Parrikar, O.
\newblock Quantum complexity of time evolution with chaotic Hamiltonians.
\newblock {\em Journal of High Energy Physics} {\bf 2020}, {\em 2020},~134.

\bibitem[Bhattacharyya \em{et~al.}(2021{\natexlab{a}})Bhattacharyya, Haque, and
  Kim]{Bhattacharyya2021}
Bhattacharyya, A.; Haque, S.S.; Kim, E.H.
\newblock Complexity from the reduced density matrix: a new diagnostic for
  chaos.
\newblock {\em Journal of High Energy Physics} {\bf 2021}, {\em 2021},~28.

\bibitem[Bhattacharyya \em{et~al.}(2021{\natexlab{b}})Bhattacharyya,
  Chemissany, Haque, Murugan, and Yan]{Arpan2021}
Bhattacharyya, A.; Chemissany, W.; Haque, S.S.; Murugan, J.; Yan, B.
\newblock {The Multi-faceted Inverted Harmonic Oscillator: Chaos and
  Complexity}.
\newblock {\em SciPost Phys. Core} {\bf 2021}, {\em 4},~2.
\newblock
  doi:{\changeurlcolor{black}\href{https://doi.org/10.21468/SciPostPhysCore.4.1.002}{\detokenize{10.21468/SciPostPhysCore.4.1.002}}}.

\bibitem[Parker \em{et~al.}(2019)Parker, Cao, Avdoshkin, Scaffidi, and
  Altman]{Parker2019}
Parker, D.E.; Cao, X.; Avdoshkin, A.; Scaffidi, T.; Altman, E.
\newblock A Universal Operator Growth Hypothesis.
\newblock {\em Phys. Rev. X} {\bf 2019}, {\em 9},~041017.
\newblock
  doi:{\changeurlcolor{black}\href{https://doi.org/10.1103/PhysRevX.9.041017}{\detokenize{10.1103/PhysRevX.9.041017}}}.

\bibitem[Dymarsky and Gorsky(2020)]{Dymarsky2020}
Dymarsky, A.; Gorsky, A.
\newblock Quantum chaos as delocalization in Krylov space.
\newblock {\em Phys. Rev. B} {\bf 2020}, {\em 102},~085137.
\newblock
  doi:{\changeurlcolor{black}\href{https://doi.org/10.1103/PhysRevB.102.085137}{\detokenize{10.1103/PhysRevB.102.085137}}}.

\bibitem[Caputa \em{et~al.}(2022)Caputa, Magan, and Patramanis]{Caputa2022}
Caputa, P.; Magan, J.M.; Patramanis, D.
\newblock Geometry of Krylov complexity.
\newblock {\em Phys. Rev. Research} {\bf 2022}, {\em 4},~013041.
\newblock
  doi:{\changeurlcolor{black}\href{https://doi.org/10.1103/PhysRevResearch.4.013041}{\detokenize{10.1103/PhysRevResearch.4.013041}}}.

\bibitem[Bertini \em{et~al.}(2021)Bertini, Heidrich-Meisner, Karrasch, Prosen,
  Steinigeweg, and \ifmmode \check{Z}\else
  \v{Z}\fi{}nidari\ifmmode~\check{c}\else \v{c}\fi{}]{Bertini2021}
Bertini, B.; Heidrich-Meisner, F.; Karrasch, C.; Prosen, T.; Steinigeweg, R.;
  \ifmmode \check{Z}\else \v{Z}\fi{}nidari\ifmmode~\check{c}\else \v{c}\fi{},
  M.
\newblock Finite-temperature transport in one-dimensional quantum lattice
  models.
\newblock {\em Rev. Mod. Phys.} {\bf 2021}, {\em 93},~025003.
\newblock
  doi:{\changeurlcolor{black}\href{https://doi.org/10.1103/RevModPhys.93.025003}{\detokenize{10.1103/RevModPhys.93.025003}}}.

\bibitem[Cvitanovic \em{et~al.}(2005)Cvitanovic, Artuso, Mainieri, Tanner,
  Vattay, Whelan, and Wirzba]{Cvitanovic2005}
Cvitanovic, P.; Artuso, R.; Mainieri, R.; Tanner, G.; Vattay, G.; Whelan, N.;
  Wirzba, A.
\newblock Chaos: classical and quantum.
\newblock {\em ChaosBook. org (Niels Bohr Institute, Copenhagen 2005)} {\bf
  2005}, {\em 69},~25.

\bibitem[Schuster and Just(2006)]{Schuster2006}
Schuster, H.G.; Just, W.
\newblock {\em Deterministic chaos: an introduction}; John Wiley \& Sons,
  2006.

\bibitem[Bohigas \em{et~al.}(1984)Bohigas, Giannoni, and Schmit]{BGS1984}
Bohigas, O.; Giannoni, M.J.; Schmit, C.
\newblock Characterization of Chaotic Quantum Spectra and Universality of Level
  Fluctuation Laws.
\newblock {\em Phys. Rev. Lett.} {\bf 1984}, {\em 52},~1--4.
\newblock
  doi:{\changeurlcolor{black}\href{https://doi.org/10.1103/PhysRevLett.52.1}{\detokenize{10.1103/PhysRevLett.52.1}}}.

\bibitem[Stöckmann(1999)]{Stockmann1999}
Stöckmann, H.J.
\newblock {\em Quantum Chaos: An Introduction}; Cambridge University Press,
  1999.
\newblock
  doi:{\changeurlcolor{black}\href{https://doi.org/10.1017/CBO9780511524622}{\detokenize{10.1017/CBO9780511524622}}}.

\bibitem[Haake(2010)]{Haake2010}
Haake, F.
\newblock {\em Quantum Signatures of Chaos}; Springer, Berlin Heidelberg,
  2010.

\bibitem[Zyczkowski(1990)]{Zyczkowski1990}
Zyczkowski, K.
\newblock Indicators of quantum chaos based on eigenvector statistics.
\newblock {\em J. Phys. A} {\bf 1990}, {\em 23},~4427.

\bibitem[Emerson \em{et~al.}(2002)Emerson, Weinstein, Lloyd, and
  Cory]{Emerson2002}
Emerson, J.; Weinstein, Y.S.; Lloyd, S.; Cory, D.G.
\newblock Fidelity Decay as an Efficient Indicator of Quantum Chaos.
\newblock {\em Phys. Rev. Lett.} {\bf 2002}, {\em 89},~284102.
\newblock
  doi:{\changeurlcolor{black}\href{https://doi.org/10.1103/PhysRevLett.89.284102}{\detokenize{10.1103/PhysRevLett.89.284102}}}.

\bibitem[Rozenbaum \em{et~al.}(2017)Rozenbaum, Ganeshan, and
  Galitski]{Rozenbaum2017}
Rozenbaum, E.B.; Ganeshan, S.; Galitski, V.
\newblock Lyapunov Exponent and Out-of-Time-Ordered Correlator's Growth Rate in
  a Chaotic System.
\newblock {\em Phys. Rev. Lett.} {\bf 2017}, {\em 118},~086801.
\newblock
  doi:{\changeurlcolor{black}\href{https://doi.org/10.1103/PhysRevLett.118.086801}{\detokenize{10.1103/PhysRevLett.118.086801}}}.

\bibitem[Garc\'{\i}a-Mata \em{et~al.}(2018)Garc\'{\i}a-Mata, Saraceno,
  Jalabert, Roncaglia, and Wisniacki]{Mata2018}
Garc\'{\i}a-Mata, I.; Saraceno, M.; Jalabert, R.A.; Roncaglia, A.J.; Wisniacki,
  D.A.
\newblock Chaos Signatures in the Short and Long Time Behavior of the
  Out-of-Time Ordered Correlator.
\newblock {\em Phys. Rev. Lett.} {\bf 2018}, {\em 121},~210601.
\newblock
  doi:{\changeurlcolor{black}\href{https://doi.org/10.1103/PhysRevLett.121.210601}{\detokenize{10.1103/PhysRevLett.121.210601}}}.

\bibitem[Chen and Ludwig(2018)]{ChenX2018}
Chen, X.; Ludwig, A.W.W.
\newblock Universal spectral correlations in the chaotic wave function and the
  development of quantum chaos.
\newblock {\em Phys. Rev. B} {\bf 2018}, {\em 98},~064309.
\newblock
  doi:{\changeurlcolor{black}\href{https://doi.org/10.1103/PhysRevB.98.064309}{\detokenize{10.1103/PhysRevB.98.064309}}}.

\bibitem[Kos \em{et~al.}(2018)Kos, Ljubotina, and Prosen]{Kos2018}
Kos, P.; Ljubotina, M.; Prosen, T.c.v.
\newblock Many-Body Quantum Chaos: Analytic Connection to Random Matrix Theory.
\newblock {\em Phys. Rev. X} {\bf 2018}, {\em 8},~021062.
\newblock
  doi:{\changeurlcolor{black}\href{https://doi.org/10.1103/PhysRevX.8.021062}{\detokenize{10.1103/PhysRevX.8.021062}}}.

\bibitem[Bertini \em{et~al.}(2018)Bertini, Kos, and Prosen]{Bertini2018}
Bertini, B.; Kos, P.; Prosen, T.c.v.
\newblock Exact Spectral Form Factor in a Minimal Model of Many-Body Quantum
  Chaos.
\newblock {\em Phys. Rev. Lett.} {\bf 2018}, {\em 121},~264101.
\newblock
  doi:{\changeurlcolor{black}\href{https://doi.org/10.1103/PhysRevLett.121.264101}{\detokenize{10.1103/PhysRevLett.121.264101}}}.

\bibitem[Gietka \em{et~al.}(2019)Gietka, Chwede\ifmmode~\acute{n}\else
  \'{n}\fi{}czuk, Wasak, and Piazza]{Gietka2019}
Gietka, K.; Chwede\ifmmode~\acute{n}\else \'{n}\fi{}czuk, J.; Wasak, T.;
  Piazza, F.
\newblock Multipartite entanglement dynamics in a regular-to-ergodic
  transition: Quantum Fisher information approach.
\newblock {\em Phys. Rev. B} {\bf 2019}, {\em 99},~064303.
\newblock
  doi:{\changeurlcolor{black}\href{https://doi.org/10.1103/PhysRevB.99.064303}{\detokenize{10.1103/PhysRevB.99.064303}}}.

\bibitem[Xu \em{et~al.}(2020)Xu, Scaffidi, and Cao]{XuT2020}
Xu, T.; Scaffidi, T.; Cao, X.
\newblock Does Scrambling Equal Chaos?
\newblock {\em Phys. Rev. Lett.} {\bf 2020}, {\em 124},~140602.
\newblock
  doi:{\changeurlcolor{black}\href{https://doi.org/10.1103/PhysRevLett.124.140602}{\detokenize{10.1103/PhysRevLett.124.140602}}}.

\bibitem[{Cao} \em{et~al.}(2021){Cao}, {Xu}, and {del Campo}]{ZanC2021}
{Cao}, Z.; {Xu}, Z.; {del Campo}, A.
\newblock {Probing quantum chaos in multipartite systems}.
\newblock {\em arXiv e-prints} {\bf 2021}, p. arXiv:2111.12475,
  \href{http://xxx.lanl.gov/abs/2111.12475}{{\normalfont
  [arXiv:quant-ph/2111.12475]}}.

\bibitem[Zonnios \em{et~al.}(2022)Zonnios, Levinsen, Parish, Pollock, and
  Modi]{Zonnios2022}
Zonnios, M.; Levinsen, J.; Parish, M.M.; Pollock, F.A.; Modi, K.
\newblock Signatures of Quantum Chaos in an Out-of-Time-Order Tensor.
\newblock {\em Phys. Rev. Lett.} {\bf 2022}, {\em 128},~150601.
\newblock
  doi:{\changeurlcolor{black}\href{https://doi.org/10.1103/PhysRevLett.128.150601}{\detokenize{10.1103/PhysRevLett.128.150601}}}.

\bibitem[{Ra{\'u}l Gonz{\'a}lez Alonso} \em{et~al.}(2022){Ra{\'u}l Gonz{\'a}lez
  Alonso}, {Shammah}, {Ahmed}, {Nori}, and {Dressel}]{Alonso2022}
{Ra{\'u}l Gonz{\'a}lez Alonso}, J.; {Shammah}, N.; {Ahmed}, S.; {Nori}, F.;
  {Dressel}, J.
\newblock {Diagnosing quantum chaos with out-of-time-ordered-correlator
  quasiprobability in the kicked-top model}.
\newblock {\em arXiv e-prints} {\bf 2022}, p. arXiv:2201.08175,
  \href{http://xxx.lanl.gov/abs/2201.08175}{{\normalfont
  [arXiv:quant-ph/2201.08175]}}.

\bibitem[Joshi \em{et~al.}(2022)Joshi, Elben, Vikram, Vermersch, Galitski, and
  Zoller]{Joshi2022}
Joshi, L.K.; Elben, A.; Vikram, A.; Vermersch, B.; Galitski, V.; Zoller, P.
\newblock Probing Many-Body Quantum Chaos with Quantum Simulators.
\newblock {\em Phys. Rev. X} {\bf 2022}, {\em 12},~011018.
\newblock
  doi:{\changeurlcolor{black}\href{https://doi.org/10.1103/PhysRevX.12.011018}{\detokenize{10.1103/PhysRevX.12.011018}}}.

\bibitem[{Lozej} \em{et~al.}(2022){Lozej}, {Lukman}, and {Robnik}]{Lozej2022}
{Lozej}, {\v{C}}.; {Lukman}, D.; {Robnik}, M.
\newblock {Phenomenology of quantum eigenstates in mixed-type systems: lemon
  billiards with complex phase space structure}.
\newblock {\em arXiv e-prints} {\bf 2022}, p. arXiv:2207.07197,
  \href{http://xxx.lanl.gov/abs/2207.07197}{{\normalfont
  [arXiv:nlin.CD/2207.07197]}}.

\bibitem[Abanin \em{et~al.}(2019)Abanin, Altman, Bloch, and Serbyn]{Abanin2019}
Abanin, D.A.; Altman, E.; Bloch, I.; Serbyn, M.
\newblock Colloquium: Many-body localization, thermalization, and entanglement.
\newblock {\em Rev. Mod. Phys.} {\bf 2019}, {\em 91},~021001.
\newblock
  doi:{\changeurlcolor{black}\href{https://doi.org/10.1103/RevModPhys.91.021001}{\detokenize{10.1103/RevModPhys.91.021001}}}.

\bibitem[Turner \em{et~al.}(2018)Turner, Michailidis, Abanin, Serbyn, and
  Papić]{Turner2018}
Turner, C.J.; Michailidis, A.A.; Abanin, D.A.; Serbyn, M.; Papić, Z.
\newblock Weak ergodicity breaking from quantum many-body scars.
\newblock {\em Nature Physics} {\bf 2018}, {\em 14},~745--749.

\bibitem[Sinha and Sinha(2020)]{Sinha2020}
Sinha, S.; Sinha, S.
\newblock Chaos and Quantum Scars in Bose-Josephson Junction Coupled to a
  Bosonic Mode.
\newblock {\em Phys. Rev. Lett.} {\bf 2020}, {\em 125},~134101.
\newblock
  doi:{\changeurlcolor{black}\href{https://doi.org/10.1103/PhysRevLett.125.134101}{\detokenize{10.1103/PhysRevLett.125.134101}}}.

\bibitem[Turner \em{et~al.}(2021)Turner, Desaules, Bull, and
  Papi\ifmmode~\acute{c}\else \'{c}\fi{}]{Turner2021}
Turner, C.J.; Desaules, J.Y.; Bull, K.; Papi\ifmmode~\acute{c}\else \'{c}\fi{},
  Z.
\newblock Correspondence Principle for Many-Body Scars in Ultracold Rydberg
  Atoms.
\newblock {\em Phys. Rev. X} {\bf 2021}, {\em 11},~021021.
\newblock
  doi:{\changeurlcolor{black}\href{https://doi.org/10.1103/PhysRevX.11.021021}{\detokenize{10.1103/PhysRevX.11.021021}}}.

\bibitem[Mondragon-Shem \em{et~al.}(2021)Mondragon-Shem, Vavilov, and
  Martin]{Shem2021}
Mondragon-Shem, I.; Vavilov, M.G.; Martin, I.
\newblock Fate of Quantum Many-Body Scars in the Presence of Disorder.
\newblock {\em PRX Quantum} {\bf 2021}, {\em 2},~030349.
\newblock
  doi:{\changeurlcolor{black}\href{https://doi.org/10.1103/PRXQuantum.2.030349}{\detokenize{10.1103/PRXQuantum.2.030349}}}.

\bibitem[Serbyn \em{et~al.}(2021)Serbyn, Abanin, and Papić]{Serbyn2021}
Serbyn, M.; Abanin, D.A.; Papić, Z.
\newblock Quantum many-body scars and weak breaking of ergodicity.
\newblock {\em Nature Physics} {\bf 2021}, {\em 17},~675--685.

\bibitem[Kloc \em{et~al.}(2017)Kloc, Stránský, and Cejnar]{Kloc2017}
Kloc, M.; Stránský, P.; Cejnar, P.
\newblock Quantum phases and entanglement properties of an extended Dicke
  model.
\newblock {\em Annals of Physics} {\bf 2017}, {\em 382},~85--111.
\newblock
  doi:{\changeurlcolor{black}\href{https://doi.org/https://doi.org/10.1016/j.aop.2017.04.005}{\detokenize{https://doi.org/10.1016/j.aop.2017.04.005}}}.

\bibitem[Rodriguez \em{et~al.}(2018)Rodriguez, Chilingaryan, and
  Rodr\'{\i}guez-Lara]{Rodriguez2018}
Rodriguez, J.P.J.; Chilingaryan, S.A.; Rodr\'{\i}guez-Lara, B.M.
\newblock Critical phenomena in an extended Dicke model.
\newblock {\em Phys. Rev. A} {\bf 2018}, {\em 98},~043805.
\newblock
  doi:{\changeurlcolor{black}\href{https://doi.org/10.1103/PhysRevA.98.043805}{\detokenize{10.1103/PhysRevA.98.043805}}}.

\bibitem[Guerra \em{et~al.}(2020)Guerra, Mahecha-Gómez, and
  Hirsch]{Guerra2020}
Guerra, C.A.E.; Mahecha-Gómez, J.; Hirsch, J.G.
\newblock Quantum phase transition and Berry phase in an extended Dicke model.
\newblock {\em The European Physical Journal D} {\bf 2020}, {\em 74},~200.

\bibitem[Herrera~Romero \em{et~al.}(2022)Herrera~Romero, Bastarrachea-Magnani,
  and Linares]{Romero2022}
Herrera~Romero, R.; Bastarrachea-Magnani, M.A.; Linares, R.
\newblock Critical Phenomena in Light-Matter Systems with Collective Matter
  Interactions.
\newblock {\em Entropy} {\bf 2022}, {\em 24}.
\newblock
  doi:{\changeurlcolor{black}\href{https://doi.org/10.3390/e24091198}{\detokenize{10.3390/e24091198}}}.

\bibitem[Dicke(1954)]{Dicke1954}
Dicke, R.H.
\newblock Coherence in Spontaneous Radiation Processes.
\newblock {\em Phys. Rev.} {\bf 1954}, {\em 93},~99--110.
\newblock
  doi:{\changeurlcolor{black}\href{https://doi.org/10.1103/PhysRev.93.99}{\detokenize{10.1103/PhysRev.93.99}}}.

\bibitem[Emary and Brandes(2003)]{Emary2003}
Emary, C.; Brandes, T.
\newblock Chaos and the quantum phase transition in the Dicke model.
\newblock {\em Phys. Rev. E} {\bf 2003}, {\em 67},~066203.
\newblock
  doi:{\changeurlcolor{black}\href{https://doi.org/10.1103/PhysRevE.67.066203}{\detokenize{10.1103/PhysRevE.67.066203}}}.

\bibitem[Wang and Robnik(2020)]{QianW2020}
Wang, Q.; Robnik, M.
\newblock Statistical properties of the localization measure of chaotic
  eigenstates in the Dicke model.
\newblock {\em Phys. Rev. E} {\bf 2020}, {\em 102},~032212.
\newblock
  doi:{\changeurlcolor{black}\href{https://doi.org/10.1103/PhysRevE.102.032212}{\detokenize{10.1103/PhysRevE.102.032212}}}.

\bibitem[Bastarrachea-Magnani \em{et~al.}(2015)Bastarrachea-Magnani, del
  Carpio, Lerma-Hern{\'{a}}ndez, and Hirsch]{Magnani2015}
Bastarrachea-Magnani, M.A.; del Carpio, B.L.; Lerma-Hern{\'{a}}ndez, S.;
  Hirsch, J.G.
\newblock Chaos in the Dicke model: quantum and semiclassical analysis.
\newblock {\em Physica Scripta} {\bf 2015}, {\em 90},~068015.
\newblock
  doi:{\changeurlcolor{black}\href{https://doi.org/10.1088/0031-8949/90/6/068015}{\detokenize{10.1088/0031-8949/90/6/068015}}}.

\bibitem[Bastarrachea-Magnani \em{et~al.}(2016)Bastarrachea-Magnani,
  L\'opez-del Carpio, Ch\'avez-Carlos, Lerma-Hern\'andez, and
  Hirsch]{Magnani2016}
Bastarrachea-Magnani, M.A.; L\'opez-del Carpio, B.; Ch\'avez-Carlos, J.;
  Lerma-Hern\'andez, S.; Hirsch, J.G.
\newblock Delocalization and quantum chaos in atom-field systems.
\newblock {\em Phys. Rev. E} {\bf 2016}, {\em 93},~022215.
\newblock
  doi:{\changeurlcolor{black}\href{https://doi.org/10.1103/PhysRevE.93.022215}{\detokenize{10.1103/PhysRevE.93.022215}}}.

\bibitem[Robles~Robles \em{et~al.}(2015)Robles~Robles, Chilingaryan,
  Rodr\'{\i}guez-Lara, and Lee]{Robles2015}
Robles~Robles, R.A.; Chilingaryan, S.A.; Rodr\'{\i}guez-Lara, B.M.; Lee, R.K.
\newblock Ground state in the finite Dicke model for interacting qubits.
\newblock {\em Phys. Rev. A} {\bf 2015}, {\em 91},~033819.
\newblock
  doi:{\changeurlcolor{black}\href{https://doi.org/10.1103/PhysRevA.91.033819}{\detokenize{10.1103/PhysRevA.91.033819}}}.

\bibitem[Cejnar \em{et~al.}(2011)Cejnar, Stránský, and Macek]{Pavel2011}
Cejnar, P.; Stránský, P.; Macek, M.
\newblock Regular and Chaotic Collective Modes in Nuclei.
\newblock {\em Nuclear Physics News} {\bf 2011}, {\em 21},~22--27,
  \href{http://xxx.lanl.gov/abs/https://doi.org/10.1080/10619127.2011.629919}{{\normalfont
  [https://doi.org/10.1080/10619127.2011.629919]}}.
\newblock
  doi:{\changeurlcolor{black}\href{https://doi.org/10.1080/10619127.2011.629919}{\detokenize{10.1080/10619127.2011.629919}}}.

\bibitem[Ch\'avez-Carlos \em{et~al.}(2019)Ch\'avez-Carlos, L\'opez-del Carpio,
  Bastarrachea-Magnani, Str\'ansk\'y, Lerma-Hern\'andez, Santos, and
  Hirsch]{Chavez2019}
Ch\'avez-Carlos, J.; L\'opez-del Carpio, B.; Bastarrachea-Magnani, M.A.;
  Str\'ansk\'y, P.; Lerma-Hern\'andez, S.; Santos, L.F.; Hirsch, J.G.
\newblock Quantum and Classical Lyapunov Exponents in Atom-Field Interaction
  Systems.
\newblock {\em Phys. Rev. Lett.} {\bf 2019}, {\em 122},~024101.
\newblock
  doi:{\changeurlcolor{black}\href{https://doi.org/10.1103/PhysRevLett.122.024101}{\detokenize{10.1103/PhysRevLett.122.024101}}}.

\bibitem[Rodr\'{\i}guez-Lara and Lee(2011)]{Lara2011}
Rodr\'{\i}guez-Lara, B.M.; Lee, R.K.
\newblock Classical dynamics of a two-species condensate driven by a quantum
  field.
\newblock {\em Phys. Rev. E} {\bf 2011}, {\em 84},~016225.
\newblock
  doi:{\changeurlcolor{black}\href{https://doi.org/10.1103/PhysRevE.84.016225}{\detokenize{10.1103/PhysRevE.84.016225}}}.

\bibitem[Berry and Tabor(1977)]{Berry1977}
Berry, M.V.; Tabor, M.
\newblock Level clustering in the regular spectrum.
\newblock {\em Proc. R. Soc. A} {\bf 1977}, {\em 356},~375--394.
\newblock
  doi:{\changeurlcolor{black}\href{https://doi.org/10.1098/rspa.1977.0140}{\detokenize{10.1098/rspa.1977.0140}}}.

\bibitem[Jacquod and Shepelyansky(1997)]{Jacquod1997}
Jacquod, P.; Shepelyansky, D.L.
\newblock Emergence of Quantum Chaos in Finite Interacting Fermi Systems.
\newblock {\em Phys. Rev. Lett.} {\bf 1997}, {\em 79},~1837--1840.
\newblock
  doi:{\changeurlcolor{black}\href{https://doi.org/10.1103/PhysRevLett.79.1837}{\detokenize{10.1103/PhysRevLett.79.1837}}}.

\bibitem[Brody \em{et~al.}(1981)Brody, Flores, French, Mello, Pandey, and
  Wong]{Brody1981}
Brody, T.A.; Flores, J.; French, J.B.; Mello, P.A.; Pandey, A.; Wong, S.S.M.
\newblock Random-matrix physics: spectrum and strength fluctuations.
\newblock {\em Rev. Mod. Phys.} {\bf 1981}, {\em 53},~385--479.
\newblock
  doi:{\changeurlcolor{black}\href{https://doi.org/10.1103/RevModPhys.53.385}{\detokenize{10.1103/RevModPhys.53.385}}}.

\bibitem[Guhr \em{et~al.}(1998)Guhr, Müller–Groeling, and
  Weidenmüller]{Guhr1998}
Guhr, T.; Müller–Groeling, A.; Weidenmüller, H.A.
\newblock Random-matrix theories in quantum physics: common concepts.
\newblock {\em Physics Reports} {\bf 1998}, {\em 299},~189--425.
\newblock
  doi:{\changeurlcolor{black}\href{https://doi.org/https://doi.org/10.1016/S0370-1573(97)00088-4}{\detokenize{https://doi.org/10.1016/S0370-1573(97)00088-4}}}.

\bibitem[Oganesyan and Huse(2007)]{Oganesyan2007}
Oganesyan, V.; Huse, D.A.
\newblock Localization of interacting fermions at high temperature.
\newblock {\em Phys. Rev. B} {\bf 2007}, {\em 75},~155111.
\newblock
  doi:{\changeurlcolor{black}\href{https://doi.org/10.1103/PhysRevB.75.155111}{\detokenize{10.1103/PhysRevB.75.155111}}}.

\bibitem[Atas \em{et~al.}(2013{\natexlab{a}})Atas, Bogomolny, Giraud, and
  Roux]{Atas2013}
Atas, Y.Y.; Bogomolny, E.; Giraud, O.; Roux, G.
\newblock Distribution of the Ratio of Consecutive Level Spacings in Random
  Matrix Ensembles.
\newblock {\em Phys. Rev. Lett.} {\bf 2013}, {\em 110},~084101.
\newblock
  doi:{\changeurlcolor{black}\href{https://doi.org/10.1103/PhysRevLett.110.084101}{\detokenize{10.1103/PhysRevLett.110.084101}}}.

\bibitem[Atas \em{et~al.}(2013{\natexlab{b}})Atas, Bogomolny, Giraud, Vivo, and
  Vivo]{Atas2013a}
Atas, Y.Y.; Bogomolny, E.; Giraud, O.; Vivo, P.; Vivo, E.
\newblock Joint probability densities of level spacing ratios in random
  matrices.
\newblock {\em Journal of Physics A: Mathematical and Theoretical} {\bf 2013},
  {\em 46},~355204.
\newblock
  doi:{\changeurlcolor{black}\href{https://doi.org/10.1088/1751-8113/46/35/355204}{\detokenize{10.1088/1751-8113/46/35/355204}}}.

\bibitem[Giraud \em{et~al.}(2022)Giraud, Mac\'e, Vernier, and Alet]{Giraud2022}
Giraud, O.; Mac\'e, N.; Vernier, E.; Alet, F.
\newblock Probing Symmetries of Quantum Many-Body Systems through Gap Ratio
  Statistics.
\newblock {\em Phys. Rev. X} {\bf 2022}, {\em 12},~011006.
\newblock
  doi:{\changeurlcolor{black}\href{https://doi.org/10.1103/PhysRevX.12.011006}{\detokenize{10.1103/PhysRevX.12.011006}}}.

\bibitem[Berry(1977)]{Berry1977a}
Berry, M.V.
\newblock Regular and irregular semiclassical wavefunctions.
\newblock {\em J. Phys. A} {\bf 1977}, {\em 10},~2083--2091.
\newblock
  doi:{\changeurlcolor{black}\href{https://doi.org/10.1088/0305-4470/10/12/016}{\detokenize{10.1088/0305-4470/10/12/016}}}.

\bibitem[Porter and Thomas(1956)]{Porter1956}
Porter, C.E.; Thomas, R.G.
\newblock Fluctuations of Nuclear Reaction Widths.
\newblock {\em Phys. Rev.} {\bf 1956}, {\em 104},~483--491.
\newblock
  doi:{\changeurlcolor{black}\href{https://doi.org/10.1103/PhysRev.104.483}{\detokenize{10.1103/PhysRev.104.483}}}.

\bibitem[Mehta(2004)]{Mehta2004}
Mehta, M.L.
\newblock {\em Random matrices}; Elsevier,  2004.

\bibitem[Izrailev(1990)]{Izrailev1990}
Izrailev, F.M.
\newblock Simple models of quantum chaos: Spectrum and eigenfunctions.
\newblock {\em Phys. Rep.} {\bf 1990}, {\em 196},~299--392.
\newblock
  doi:{\changeurlcolor{black}\href{https://doi.org/https://doi.org/10.1016/0370-1573(90)90067-C}{\detokenize{https://doi.org/10.1016/0370-1573(90)90067-C}}}.

\bibitem[Haake and \ifmmode~\dot{Z}\else \.{Z}\fi{}yczkowski(1990)]{HaakeZ1990}
Haake, F.; \ifmmode~\dot{Z}\else \.{Z}\fi{}yczkowski, K.
\newblock Random-matrix theory and eigenmodes of dynamical systems.
\newblock {\em Phys. Rev. A} {\bf 1990}, {\em 42},~1013--1016.
\newblock
  doi:{\changeurlcolor{black}\href{https://doi.org/10.1103/PhysRevA.42.1013}{\detokenize{10.1103/PhysRevA.42.1013}}}.

\bibitem[Leboeuf and Voros(1990)]{Leboeuf1990a}
Leboeuf, P.; Voros, A.
\newblock Chaos-revealing multiplicative representation of quantum eigenstates.
\newblock {\em J. Phys. A: Mathematical and General} {\bf 1990}, {\em
  23},~1765--1774.
\newblock
  doi:{\changeurlcolor{black}\href{https://doi.org/10.1088/0305-4470/23/10/017}{\detokenize{10.1088/0305-4470/23/10/017}}}.

\bibitem[Wang and Robnik(2021)]{QianWg2021}
Wang, Q.; Robnik, M.
\newblock Multifractality in Quasienergy Space of Coherent States as a
  Signature of Quantum Chaos.
\newblock {\em Entropy} {\bf 2021}, {\em 23}.
\newblock
  doi:{\changeurlcolor{black}\href{https://doi.org/10.3390/e23101347}{\detokenize{10.3390/e23101347}}}.

\bibitem[Torres-Herrera \em{et~al.}(2016)Torres-Herrera, Karp, Távora, and
  Santos]{Herrera2016}
Torres-Herrera, E.J.; Karp, J.; Távora, M.; Santos, L.F.
\newblock Realistic Many-Body Quantum Systems vs. Full Random Matrices: Static
  and Dynamical Properties.
\newblock {\em Entropy} {\bf 2016}, {\em 18}.
\newblock
  doi:{\changeurlcolor{black}\href{https://doi.org/10.3390/e18100359}{\detokenize{10.3390/e18100359}}}.

\bibitem[B\"acker \em{et~al.}(2019)B\"acker, Haque, and Khaymovich]{Backer2019}
B\"acker, A.; Haque, M.; Khaymovich, I.M.
\newblock Multifractal dimensions for random matrices, chaotic quantum maps,
  and many-body systems.
\newblock {\em Phys. Rev. E} {\bf 2019}, {\em 100},~032117.
\newblock
  doi:{\changeurlcolor{black}\href{https://doi.org/10.1103/PhysRevE.100.032117}{\detokenize{10.1103/PhysRevE.100.032117}}}.

\bibitem[{Nakerst} and {Haque}(2022)]{Nakerst2022}
{Nakerst}, G.; {Haque}, M.
\newblock {Chaos in the three-site Bose-Hubbard model -- classical vs quantum}.
\newblock {\em arXiv e-prints} {\bf 2022}, p. arXiv:2203.09953,
  \href{http://xxx.lanl.gov/abs/2203.09953}{{\normalfont
  [arXiv:quant-ph/2203.09953]}}.

\bibitem[Khaymovich \em{et~al.}(2019)Khaymovich, Haque, and
  McClarty]{Khaymovich2019}
Khaymovich, I.M.; Haque, M.; McClarty, P.A.
\newblock Eigenstate Thermalization, Random Matrix Theory, and Behemoths.
\newblock {\em Phys. Rev. Lett.} {\bf 2019}, {\em 122},~070601.
\newblock
  doi:{\changeurlcolor{black}\href{https://doi.org/10.1103/PhysRevLett.122.070601}{\detokenize{10.1103/PhysRevLett.122.070601}}}.

\bibitem[Pausch \em{et~al.}(2021)Pausch, Carnio, Rodr\'{\i}guez, and
  Buchleitner]{Pausch2021}
Pausch, L.; Carnio, E.G.; Rodr\'{\i}guez, A.; Buchleitner, A.
\newblock Chaos and Ergodicity across the Energy Spectrum of Interacting
  Bosons.
\newblock {\em Phys. Rev. Lett.} {\bf 2021}, {\em 126},~150601.
\newblock
  doi:{\changeurlcolor{black}\href{https://doi.org/10.1103/PhysRevLett.126.150601}{\detokenize{10.1103/PhysRevLett.126.150601}}}.

\bibitem[Beugeling \em{et~al.}(2018)Beugeling, B\"acker, Moessner, and
  Haque]{Wouter2018}
Beugeling, W.; B\"acker, A.; Moessner, R.; Haque, M.
\newblock Statistical properties of eigenstate amplitudes in complex quantum
  systems.
\newblock {\em Phys. Rev. E} {\bf 2018}, {\em 98},~022204.
\newblock
  doi:{\changeurlcolor{black}\href{https://doi.org/10.1103/PhysRevE.98.022204}{\detokenize{10.1103/PhysRevE.98.022204}}}.

\bibitem[Luitz and Bar~Lev(2016)]{Luitz2016}
Luitz, D.J.; Bar~Lev, Y.
\newblock Anomalous Thermalization in Ergodic Systems.
\newblock {\em Phys. Rev. Lett.} {\bf 2016}, {\em 117},~170404.
\newblock
  doi:{\changeurlcolor{black}\href{https://doi.org/10.1103/PhysRevLett.117.170404}{\detokenize{10.1103/PhysRevLett.117.170404}}}.

\bibitem[Luitz \em{et~al.}(2020)Luitz, Khaymovich, and Lev]{Luitz2020}
Luitz, D.J.; Khaymovich, I.M.; Lev, Y.B.
\newblock {Multifractality and its role in anomalous transport in the
  disordered XXZ spin-chain}.
\newblock {\em SciPost Phys. Core} {\bf 2020}, {\em 2},~6.
\newblock
  doi:{\changeurlcolor{black}\href{https://doi.org/10.21468/SciPostPhysCore.2.2.006}{\detokenize{10.21468/SciPostPhysCore.2.2.006}}}.

\bibitem[Kullback and Leibler(1951)]{Kullback1951}
Kullback, S.; Leibler, R.A.
\newblock On Information and Sufficiency.
\newblock {\em Ann. Math. Stat.} {\bf 1951}, {\em 22},~79--86.

\bibitem[Frisch \em{et~al.}(2014)Frisch, Mark, Aikawa, Ferlaino, Bohn,
  Makrides, Petrov, and Kotochigova]{Frisch2014}
Frisch, A.; Mark, M.; Aikawa, K.; Ferlaino, F.; Bohn, J.L.; Makrides, C.;
  Petrov, A.; Kotochigova, S.
\newblock Quantum chaos in ultracold collisions of gas-phase erbium atoms.
\newblock {\em Nature} {\bf 2014}, {\em 507},~475--479.

\end{thebibliography}

\end{document}